\documentclass[secthm,aoas,nameyear,seceqn,dvips]{arximspdf}
\usepackage{graphics}

\doi{10.1214/10-AOAS332}
\volume{4}
\issue{3}
\pubyear{2010}
\firstpage{1498}
\lastpage{1516}

\makeatletter

\newtheorem{theorem}{Theorem}[section]

\makeatother

\begin{document}
\begin{frontmatter}

\title{Variable selection and regression analysis for
graph-structured covariates with\break an application to genomics\thanksref{T1}}
\runtitle{Variable Selection for Graph Covariates}
\thankstext{T1}{Supported in part by NIH Grants CA127334 and ES009111.}

\begin{aug}
\author{\fnms{Caiyan} \snm{Li}\ead[label=e1]{licaiyan@mail.med.upenn.edu}\corref{}}
\and
\author{\fnms{Hongzhe} \snm{Li}\ead[label=e2]{hongzhe@upenn.edu}
\ead[label=u1,url]{http://www.cceb.med.upenn.edu/\textasciitilde hli}}
\runauthor{C. Li and H. Li}
\affiliation{University of Pennsylvania School of Medicine}
\address{Department of Biostatistics and Epidemiology\\
University of Pennsylvania School of Medicine\\
Philadelphia, Pennsylvania 19104\\USA\\
\printead{e1}\\
\phantom{E-mail: }\printead*{e2}}
\end{aug}

\received{\smonth{6} \syear{2009}}
\revised{\smonth{10} \syear{2009}}

%
\begin{abstract}
Graphs and networks are common ways of depicting biological
information. In biology, many different biological processes are
represented by graphs, such as regulatory networks, metabolic
pathways and protein--protein interaction networks. This kind of a
priori use of graphs is a useful supplement to the standard
numerical data such as microarray gene expression data. In this
paper we consider the problem of {regression} analysis and variable
selection when the covariates are linked on a graph. We study a
graph-constrained regularization procedure and its theoretical
properties for \mbox{regression} analysis to take into account the
neighborhood information of the variables measured on a graph. This
procedure involves a smoothness penalty on the coefficients that is
defined as a quadratic form of the Laplacian matrix associated with
the graph. We establish estimation and model selection consistency
results and provide estimation bounds for both fixed and diverging
numbers of parameters in regression models. We demonstrate by
simulations and a real data set that the proposed procedure can lead
to better variable selection and prediction than existing methods
that ignore the graph information associated with the covariates.
\end{abstract}

%

\end{frontmatter}

\section{Introduction}

There has been a growing interest in penalized least squares
problems via $L_1$ or other types of regularization, especially in
high-dimensional settings. Important penalty functions that can
lead to sparse variable selection in regression include Lasso
[Tibshirani (\citeyear{1996Tibshirani})] and SCAD [Fan and Li (\citeyear{2001Fan})]. In particular,
Lasso has the crucial advantage of being a convex problem, which
leads to efficient computational algorithms by coordinate descent
[Efron et~al. (\citeyear{2004Efron}); Friedman et~al. (\citeyear{2007Friedman}); Wu and
Lange (\citeyear{2008Wu})] and sparse solutions. Zou (\citeyear{2006Zou}) proposed a novel
adaptive Lasso procedure and presented results on model selection
consistency and oracle properties of the parameter estimates. Zhao
and Yu (\citeyear{2006Zhao}) presented the irrepresentable condition for model
selection consistency of Lasso. Zhang and Huang (\citeyear{2006Zhang}) studied the
sparsity and bias of the Lasso selection in high-dimensional linear
regression. Fan and Li (\citeyear{2001Fan}) and Huang and Xie (\citeyear{2007Huang}) established
the asymptotic oracle properties of the SCAD-penalized least squares
estimators when the number of covariates is fixed or increases with
the sample sizes. These novel penalized estimation methods are quite
effective in selecting relevant variables and in predicting future
outcomes, especially in high-dimensional settings.

New estimation procedures have also been developed in recent years to
account for certain structures of the
explanatory variables. These include the group Lasso procedure [Yuan
and Lin (\citeyear{2006Yuan})]
when the explanatory variables are grouped
or organized in a hierarchical manner, the elastic
net (Enet) procedure [Zou and Hastie (\citeyear{2005Zou})] that deals with groups of
highly correlated variables, and the
fused Lasso [Tibshirani et~al. (\citeyear{2005Tibshirani})] that imposes
the $L_1$ penalty on the absolute
differences of the regression coefficients in order to account for some
smoothness of the coefficients.
Nardi and Rinaldo (\citeyear{2008Nardi}) established the asymptotic properties of the
group Lasso estimator for linear models.
Jia and Yu (\citeyear{2008Jia}) provided conditions for model selection consistency
of the elastic net when $p\gg n$.
Zou and Zhang (\citeyear{2009Zou}) proposed an adaptive elastic net with a diverging
number of parameters and established
its oracle property.
Among these procedures, the Enet regularization and the fused Lasso
are particularly appropriate for the analysis of genomic data, where
the former encourages a grouping effect and the latter often leads
to smoothness of the coefficient profiles for ordered covariates.

Motivated by a genomic application in order to account for network
information in the analysis of genomic data, Li and Li (\citeyear{2008Li})
proposed a network-constrained regularization procedure for fitting
linear regression models and for variable selection, where the
predictors in the regression model are genomic data that are
measured on the genetic networks, which we call the
graph-structured covariates. In particular, we assume that the
covariates in the regression model are values of the nodes on a
graph, where a link between two nodes may indicate a functional
relationship between two genes in a genetic network or physical
neighborhood between two voxels on brain images. Since many
biological networks are constructed using data from high-throughput
experiments, often there is a probability associated with an edge to
indicate the certainty of a link. Such an edge probability can be
used as a weight in a undirected graph, in which case we have a
weighted graph. This graph-constrained regularization procedure is
similar in spirit to the fused Lasso [Tibshirani et~al.
(\citeyear{2005Tibshirani})], both of which try to smooth the regression coefficients in
certain ways. However, the fused Lasso does not utilize prior graph
information. Second, instead of using the $L_2$ norm on the
differences of the coefficients of the linked variables, the fused
Lasso uses the $L_1$ norm on the differences, which tends to lead to
the same regression coefficients for linked variables. In this
paper we consider the general problem of regression analysis when
the explanatory variables are nodes on a graph and present
a cyclical coordinate descent algorithm [Friedman et~al. (\citeyear{2007Friedman})] to implement the network-constrained
regularization procedure of Li and Li (\citeyear{2008Li}). This algorithm provides
new insight on how
neighboring variables affect the coefficient estimate of a node. We
also extend the
procedure of Li and Li (\citeyear{2008Li}) to account for the possibility of
different signs of the regression coefficients for neighboring
variables. In addition, we provide theoretical results of the
estimates, including sign consistency and
error bounds of the estimator and $L_2$ consistency.

This paper is organized as follows. In Section \ref{reg} we
describe the problem of regression analysis with covariates measured
on graphs. We then present a graph-constrained estimation (Grace)
procedure in order to account for the graph structures in Section
\ref{net} and an efficient coordinate descent algorithm to implement
the proposed regularization methods in Section \ref{algo}. We
present the estimation and model selection consistency results in
Section \ref{theory}. We provide Monte Carlo simulation results in
Section \ref{simu} and results from the application to the analysis
of a data set on the gene expression of brain aging in Section
\ref{realdata}. Finally, we give a brief discussion of the methods
and results.

\section{Regression analysis for covariates measured on a graph}
\label{reg}

Consider a weighted graph $G=(V,E, W)$, where $V=\{1,\ldots, p\}$ is
the set of vertices that correspond to the $p$ predictors,
$E=\{u\sim v\}$ is the set of edges indicating that the predictors
$u$ and $v$ are linked on the graph and there is an edge
between $u$ and $v$, and $W$ is the set of
weights of the edges, where $w(u,v)$ denotes the weight of edge
$e=(u\sim v)$. In genomic studies, biological networks are often
represented
as graphs, an edge between $u$ and $v$ on the graph
can indicate some functional relationship between them and the weight
can be used to measure
the uncertainty
of the edge between two vertices, for example, indicating the
probability of having an edge between two variables when the graph
is constructed from data. For each given sample, we assume that we
have numerical measurements on each of the vertices and these
measurements are treated as explanatory variables in a regression
analysis framework. For the $u$th node, let $x_{iu}$ be the
numerical measurement of the $u$th vertex on the graph for the $i$th
individual. Further, let $\mathrm{x}_u=(x_{1u},\ldots,x_{nu})^T$ be
the measured values at the $u$th vertex for $n$ i.i.d. samples.
Consider the problem of variable selection and estimation where we
have design matrix $\mathrm{X}=(\mathrm x_1,\mathrm x_2,\ldots,\mathrm
x_{p})\in
\mathcal{R}^{n\times p}$ and response vector $\mathrm{y}=(y_1, y_2,
\ldots,
y_n)^T\in\mathcal{R}^n$, and they follow a linear model
%
\begin{equation}\label{true}
\textrm{y}=\textrm{X}\beta+\varepsilon,
\end{equation}
where $\varepsilon=(\varepsilon_1,\ldots,\varepsilon_n)^T \sim N(0,
\sigma^2 I_n)$ and $\beta=(\beta_1,\ldots,\beta_{p})^T$. Throughout
this paper we assume that the predictors and the response are
centered so that
\begin{eqnarray*}
\sum_{i=1}^{n}y_i=0,\qquad \sum_{i=1}^{n}x_{ij}=0, \quad\mbox{and}\quad \frac{1
}{ n}\sum_{i=1}^{n}x_{ij}^2=1
\qquad\mbox{for } j=1,\ldots,p.
\end{eqnarray*}
In this paper we consider that the design matrix $\mathrm{X}$ is a
deterministic matrix in the fixed design settings.

When $p$ is large, we assume that model (\ref{true}) is ``sparse,''
that is, most of the true regression coefficients $\beta$ are exactly
zero. Without loss of generality, we assume the first $q$ elements of
vector $\beta$ are nonzeroes. Denote
$\beta_{(1)}=(\beta_1,\ldots,\beta_{q})^T$ and
$\beta_{(2)}=(\beta_{q+1},\ldots,\beta_{p})^T$, then element-wise
$\beta_{(1)}\neq0$ and $\beta_{(2)}=0$. Now write $\mathrm
{X}_{(1)}$ and
$\mathrm{X}_{(2)}$ as the first $q$ and last $p-q$ columns of $\mathrm{X}$,
respectively, and let $\mathrm{C}=\frac{1 }{ n}{\mathrm{X}}^T \mathrm
{X}$, which can then
be expressed in the following block-wise form:
\begin{eqnarray*}
{\mathrm{C}}= \pmatrix{
\mathrm{C}_{11} & \mathrm{C}_{12}
\cr
\mathrm{C}_{21} & \mathrm{C}_{22}
}.
\end{eqnarray*}
The goal of this paper is to develop a regularization procedure for
selecting the true
relevant variables. Different from the existing approaches, we
particularly account for
the fact that the explanatory variables are related on a graph. We make
this more precise in
the next section.

\subsection{Graph-constrained regularization and variable
selection}\label{net}

In order to account for the fact that the $p$ explanatory variables
are measured on a graph, we first introduce the Laplacian matrix
[Chung (\citeyear{1997Chung})] associated with a graph. Let the degree of the vertex
$v$ be $d_v=\sum_{u\sim v}{w(u,v)}$. We say $u$ is an isolated
vertex if $d_u=0$. Let $w(u,u)=0$. Following Chung (\citeyear{1997Chung}), we define the Laplacian
matrix~$\mathrm{L}$ for graph $G$ with the $uv$th element defined by
%
\begin{equation}\label{Lap}
\mathrm{L}(u,v)=
\cases{
1-w(u,u)/d_u, & \quad\mbox{if} $u=v$ and $d_u \neq0$,
\cr
-{w(u,v) / \sqrt{d_ud_v}}, & \quad\mbox{if} $u$ and $v$ are adjacent,
\cr
0, &
\quad\mbox{otherwise}.
}
\end{equation}
It is easy to verify that this matrix is positive semi-definite with
0 as the smallest eigenvalue and 2 as the largest eigenvalue when
all the weights are 1 [Chung (\citeyear{1997Chung})]. To allow the matrix to change
with $n$, we further express this matrix in block-wise form,
\begin{eqnarray*}
\mathrm{L}= \pmatrix{
\mathrm{L}_{11} & \mathrm{L}_{12}
\cr
\mathrm{L}_{21} & \mathrm{L}_{22}
},
\end{eqnarray*}
where $\mathrm{L}_{11}$ corresponds to the $q$ nodes that are relevant to
the response and $\mathrm{L}_{22}$ corresponds to the $p-q$ nodes that are
not relevant.

The Laplacian matrix has the following interpretations. For a given
vector $\beta$, the edge derivative of $\beta$ along the edge
$e(u,v)$ at $u$ is defined as
\[
\frac{\partial\beta}{\partial e}\bigg\vert_u=\sqrt{w(u,v)}
\biggl(\frac{\beta_u }{\sqrt{d_u}}-\frac{\beta_v }{
\sqrt{d_v}}\biggr),
\]
and, therefore, the local variation of $\beta$
at $u$ can be measured by
\[
\sqrt{\sum\biggl(\frac{\partial\beta}{\partial e}\bigg\vert_u
\biggr)^2}.
\]
The smoothness of vector $\beta$ with respect to the graph structure
can be expressed as
\[
\beta^T \mathrm{L}\beta=
\sum_{u\sim v}\biggl(\frac{\beta_u }{\sqrt{d_u}}-\frac{\beta_v }{
\sqrt{d_v}}\biggr)^{2}w(u,v).
\]
%
This variation functional for vectors $\beta$ penalizes vectors that
differ too much over nodes that are linked. It
contains a scaling by $\sqrt{d_u}$. One intuitive reason for such a
scaling is to allow a small number of nodes
with large $d_u$ to have more extreme values of $\beta_u$ while the
usually much greater number of nodes with small
$d_u$ should not ordinarily allow to have very large $\beta_u$. This
variation functional has been widely used
in semi-supervised learning on graphs [Zhu (\citeyear{2005Zhu}); Zhou et~al. (\citeyear{2004Zhou})].

For many problems with covariates measured on a graph, we would
expect that the neighboring variables are correlated and, therefore,
the regression coefficients would show some smoothness. One way to
account for such a dependence of the regression coefficients is to
impose a Markov random field (MRF) prior to the collection of
$\beta$ vectors. The MRF decomposes the joint prior distribution of
the $\beta_u$'s into lower-dimensional distributions based on the
graph-neighborhood structures. A common MRF model is the Gaussian
MRF model that assumes that the joint distribution of $\beta$ is
given by
\[
f(\beta) \propto\exp\biggl\{-\frac{1}{2\sigma^2} \beta^T \mathrm{L}
\beta\biggr\},
\]
which is an improper density.
Based on this Gaussian MRF prior assumption,
Li and Li (\citeyear{2008Li}) introduced the following graph-constrained estimation
of the regression coefficients, denoted by $\hat{\beta}$,
%
\begin{equation}\label{Net}
\hat{\beta}=\operatorname{argmin}\limits_{\beta}  Q(\beta,\lambda_1, \lambda_2),
\end{equation}
where
\begin{eqnarray}
Q(\beta,\lambda_1,
\lambda_2)&=&\|\mathrm{y}-\mathrm{X}\beta\|^2_2+\lambda_1\|\beta
\|_1+\lambda_2
\beta^T \mathrm{L}\beta\nonumber
\\
&=&(\mathrm{y}-\mathrm
{X}\beta)^T
(\mathrm{y}-\mathrm{X}\beta)+\lambda_1 \sum_u|\beta_u|\nonumber
\\
&&{} +\lambda_2 \sum_{u\sim v}\biggl(\frac{\beta_u}{
\sqrt{d_u}}-\frac{\beta_v }{\sqrt{d_v}}\biggr)^{2}w(u,v),\nonumber
\end{eqnarray}
where $\mathrm{L}$ is the Laplacian as defined in (\ref{Lap}) and the
tuning parameters $\lambda_1,\lambda_2$ control the amount of
regularization for sparsity and smoothness. For the special case
when $\lambda_2=0$, the estimate reduces to the Lasso, and when
$\mathrm{L}$ is the identity matrix, the estimate reduces to the elastic
net estimates.

\subsection{An adaptive graph-constrained regularization}

The Grace procedure may not perform well when two variables that are
linked on the graph have different signs in their regression
coefficients, in which case the coefficients are not expected to be
locally smooth. For example, for genetic networks, two genes might
be negatively correlated with the phenotypes and are therefore
expected to have different signs in their regression coefficients.
In order to account for the sign differences, we can first perform
a standard least square regression when $p<n$ or the elastic net
regression when $p\ge n$ and denote the estimate as $\tilde{\beta}$.
We can then modify the above objective function as
\begin{eqnarray*}\label{adptNet}
Q^*(\lambda_1, \lambda_2,
\beta)&=&\|\mathrm{y}-\mathrm{X}\beta\|^2_2+\lambda_1\|\beta
\|_1+\lambda_2
\beta^T
\mathrm{L}^{*} \beta,
\\ \nonumber
&=&\|\mathrm{y}-\mathrm{X}\beta\|^2_2 + \lambda_1 \sum_{j=1}^p
|\beta_j|
\\
&&{} + {\lambda_2\sum_{u\sim
v}\biggl(\frac{\mathrm{sign}(\tilde{\beta}_u)\beta_u }{
\sqrt{d_u}}-\frac{\operatorname{sign}(\tilde{\beta}_v)\beta_v }{
\sqrt{d_v}}\biggr)^{2}w(u,v)},
\end{eqnarray*}
where
\begin{eqnarray*}\label{Lap1} \mathrm{L}^{*}(u,v)=
\cases{
1-w(u,u)/d_u, &\quad \mbox{if} $u=v$ and $d_u \neq0$,
\cr
-\operatorname{sign}(\tilde{\beta}_u)
\operatorname{sign}(\tilde{\beta}_v){w(u,v) / \sqrt{d_ud_v}}, &
\quad\mbox{if} $u$ and $v$ are adjacent,
\cr
0, &
\quad\mbox{otherwise}.
}
\end{eqnarray*}
Note that the $\mathrm{L}^{*}$ matrix is still positive semi-definite. We
call the $\beta$ defined by
%
\begin{eqnarray}\label{adaptnet}
\hat{\beta}=\operatorname{argmin}\limits_{\beta}  Q^{*}(\beta,\lambda_1,
\lambda_2)
\end{eqnarray}
the adaptive Grace (aGrace).

\subsection{A coordinate descent algorithm}\label{algo}

Friedman et~al. (\citeyear{2007Friedman}) presented a coordinate descent
algorithm for solving the Lasso and the Enet regularization. In this
section we develop a similar algorithm for the proposed
\mbox{graph-constrained} regularization. We only present the detailed
algorithm for the optimization problem defined by equation
(\ref{Net}). Similar algorithms can be developed by the aGrace
defined by (\ref{adaptnet}).
If we let $\lambda=(\lambda_1+2\lambda_2)/2n$ and
$\alpha=\lambda_1/(\lambda_1+2\lambda_2)$, the Grace can be written
as
%
\begin{equation}\label{Network}
\hat{\beta}(\lambda,
\alpha)=\operatorname{argmin}\limits_{\beta} \biggl\{R(\beta):=\frac{1 }{
2n}\|\mathrm{y}-\mathrm{X}\beta\|_2^2+\lambda P_{\alpha}(\beta
)\biggr\},
\end{equation}
where
\begin{eqnarray*}
P_{\alpha}(\beta):=(1-\alpha)\frac{1 }{2} \beta^T \mathrm{L}\beta
+\alpha
\|\beta\|_1=(1-\alpha)\frac{1 }{2}\sum_{u \sim v}\biggl(\frac{\beta_u
}{
\sqrt{d_u}}-\frac{\beta_v }{\sqrt{d_v}}\biggr)^2+\alpha
\sum_{u=1}^p|\beta_u|
\end{eqnarray*}
is the graph-constrained penalty function.

Given a covariate $\mathrm x_u$, suppose we have estimated
$\tilde{\beta}_v$ for $v\neq u$, and we want to partially minimize
the objective function with respect to $\beta_u$. We can rewrite the
objective function in $(\ref{Network})$ as
\begin{eqnarray*}
R(\beta)&=&\frac{1}{2n} \sum_{i=1} \biggl(y_i-\sum_{v \neq
u}x_{iv}\tilde{\beta}_v-x_{iu}\beta_u\biggr)^2+ \lambda(1-\alpha)\frac{1
}{
2}\sum_{v \sim u}\biggl(\frac{\beta_u }{\sqrt{d_u}}-\frac{\tilde{\beta}_v
}{
\sqrt{d_v}}\biggr)^2
\\
&&{}+\lambda\alpha|\beta_u|+ \lambda(1-\alpha)\frac{1 }{
2}\mathop{\sum_{w \sim v}}_{w,v \neq u}\biggl(\frac{\tilde{\beta}_w
}{
\sqrt{d_w}}-\frac{\tilde{\beta}_v }{
\sqrt{d_v}}\biggr)^2+\lambda\alpha\sum_{w\neq u}|\tilde{\beta}_w|.
\end{eqnarray*}
We would like to compute the gradient at $\beta_u$, which only
exists when $\beta_u \neq0$. We first consider the case that the
covariate $u$ is connected to some other nodes (variables) on the
network. If $\beta_u>0$, due to the standardization of the
covariates, we can differentiate the objective function $R(\beta)$
with respect to $\beta_u$ as
\begin{eqnarray*}
\frac{\partial R }{\partial\beta_u}
&=&-\biggl[\frac{1}{ n}\sum_{i=1}
x_{iu}\biggl(y_i-\sum_{v \neq
u}x_{iv}\tilde{\beta}_v\biggr)+\lambda(1-\alpha)\sum_{v \sim
u}\frac{\tilde{\beta}_v }{\sqrt{d_ud_v}}\biggr]
\\
&&{}+\lambda
\alpha+[1+\lambda(1-\alpha)]\beta_u .
\end{eqnarray*}
Similarly, we can get the corresponding expression when $\beta_u<0$.
Following the calculus by Donoho and Johnstone (\citeyear{1994Donoho})
and Friedman et~al. (\citeyear{2007Friedman}), we obtain the coordinate-wise
update form for $\beta_u$ as
%
\begin{equation}\label{connected}
\qquad \tilde{\beta}_u \leftarrow\frac{S((1 /n)\sum_{i=1}
x_{iu}(y_i-\tilde{y}_i^{(u)})+\lambda(1-\alpha)\sum_{v \sim
u}{(\tilde{\beta}_v /\sqrt{d_ud_v})},\lambda\alpha)
}{
1+\lambda(1-\alpha)},
\end{equation}
where:
\begin{itemize}[$\bullet$]
\item[$\bullet$] $\tilde{y}_i^{(u)}=\sum_{v \neq
u}x_{iv}\tilde{\beta}_v$ is the partial residual for fitting
$\beta_u$, that is, the fitted value excluding the contribution from
$x_{iu}$. Since the covariates are standardized, $\frac{1 }{
n}\sum_{i=1} x_{iu}(y_i-\sum_{v \neq u}x_{iv}\tilde{\beta}_v)$ is
the simple least-squares coefficient while fitting the partial
residual to $x_{iu}$, $i=1,\ldots,n$.

\item[$\bullet$] $S(z,\gamma)$ is the soft-thresholding operator
with value
\begin{eqnarray*}
\operatorname{sign}(z)(|z|-\gamma)_+= \cases{
z-\gamma,& \quad\mbox{if} $z>0$ and $\gamma<|z|$,
\cr
z+\gamma,& \quad\mbox{if} $z<0$ and $\gamma<|z|$,
\cr
0, & \quad\mbox{otherwise}.
}
\end{eqnarray*}
\end{itemize}

 When covariate $u$ is not connected to other nodes on the
network, that is, when it has no neighbors, the corresponding
coordinate-wise updating formula becomes the Lasso updating
formula, that is
%
\begin{eqnarray}\label{unconnected}
\tilde{\beta}_u \leftarrow S\biggl(\frac{1 }{ n}\sum_{i=1}
x_{iu}\bigl(y_i-\tilde{y}_i^{(u)}\bigr),\lambda\alpha\biggr) .
\end{eqnarray}

 Comparing the two updated forms of (\ref{connected}) and
$(\ref{unconnected})$, an intuitive explanation can be drawn to help
to understand the effect of the graph-constraint penalty on the
coefficients. For an isolated predictor, the graph penalty is
vanished and, thus, we only apply a soft-thresholding for the Lasso
penalty, while for a connected predictor, form $(\ref{connected})$
takes into account the graph-constraint to the penalty by adding the
scaled summation of the coefficients of the neighboring covariates
to the simple least-squares coefficient and applying a proportional
shrinkage for the graph penalty.

 Given $\alpha$, we can compute the solution path for a
decreasing sequence of values for $\lambda$, starting from the
smallest value $\lambda_{\mathrm{max}}$ for which there is no
covariate selected, that is, $\hat{\beta}=0$. Similar to Friedman
et~al. (\citeyear{2007Friedman}), we can set
$\lambda_{\mathrm{max}}=\mathrm{max}_l|\langle x_l,y\rangle |/n\alpha$,
$\lambda_{\mathrm{min}}=\epsilon\lambda_{\mathrm{max}}$ and
construct a sequence of~$K$ values of~$\lambda$ decreasing from
$\lambda_{\mathrm{max}}$ to $\lambda_{\mathrm{min}}$ on the log
scale. Typical values are $\epsilon=0.001$ and $K=100$.
Cross-validation (CV) can be used to select the two tuning
parameters $\alpha$ and $\lambda$.

\section{Error bound and model selection consistency for fixed and
diverging~$p$}\label{theory}

In this section we provide some theoretical results on the proposed
Grace procedure, including the error bounds, $L_2$ consistency of
Grace and the model selection consistency for both fixed and
diverging $p$ when $p$ tends to infinity with the sample size $n$.
Our theoretical development follows that of Zhao and Yu (\citeyear{2006Zhao}), Jia
and Yu (\citeyear{2008Jia}) and Zou and Zhang (\citeyear{2009Zou}) on sign consistency of Lasso
and adaptive elastic net estimates.
In our theoretical analysis, we assume the following regularity
conditions throughout:
\begin{longlist}[(A1)]
\item[(A1)] We use $\Lambda_{\mathrm{min}}(\mathrm{M})$ and
$\Lambda_{\mathrm{max}}(\mathrm{M})$ to denote the minimum and maximum
eigenvalues of a positive definite matrix $\mathrm{M}$, respectively. We
further assume that $\mathrm{C}=\frac{1 }{ n}\mathrm{X}^T \mathrm{X}$
is positive definite
and
\[
b\le\Lambda_{\mathrm{min}}(\mathrm{C}) \le\Lambda_{\mathrm
{max}}(\mathrm{C})
\le B,
\]
where
$b$ and $B$ are two positive constants that do not depend on $n$.

\item[(A2)] $\frac{1 }{ n}\max_{1\leq i \leq n}\sum_{j=1}^p
x_{ij}^2\rightarrow0$, as $n\rightarrow\infty$.
\end{longlist}

These two conditions assume that the predictor matrix has a
reasonably good behavior and were also assumed in Zhao and Yu (\citeyear{2006Zhao})
and in Zou and Zhang (\citeyear{2009Zou}). Condition (A1) is also the condition
(F) in Fan and Peng (\citeyear{2004Fan}) and condition (A2) ensures that the rows
of the matrix $\mathrm{X}$ behave like a sample from a probability
distribution in $\mathcal{R}^p$ [Portnoy (\citeyear{1984Portnoy})]. These two
conditions hold naturally if one assumes that $x_i$ are i.i.d. with
finite second moments.

\subsection{Error bound and $L_2$-consistency of Grace}
We first provide the following nonasymptotic risk bound for the
Grace of the regression coefficients defined by~(\ref{net}) for any
$p$ and $n$:

\begin{theorem}\label{thm3.1}
Given the data $(\mathrm{y},\mathrm{X})$, define the Grace as
\begin{eqnarray*}
\hat{\beta}(\lambda_1,\lambda_2)=\operatorname{argmin}\limits_{\beta}
\{\|\mathrm{y}-\mathrm{X}\beta\|^2_2+\lambda_1\|\beta\|_1+\lambda
_2 \beta^T
\mathrm{L}
\beta\},
\end{eqnarray*}
for nonnegative tuning parameters $\lambda_1$ and $\lambda_2$. Then
under the regularity condition~\textup{({A1})}, we have
%
\begin{equation}\label{bounds}
E(\|\hat{\beta}(\lambda_1,\lambda_2)-\beta\|_2^2)\leq
\frac
{4\lambda_2^2\Lambda_{\max}^2(\mathrm{L})\|\beta_{(1)}\|_2^2+4pn
B\sigma^2+2\lambda_1^2 p}{n^2
\Lambda_{\min}^2(\mathrm{C}+(\lambda_2/n) \mathrm{L})}.
\end{equation}
\end{theorem}

The proof of this theorem is given in Li and Li (\citeyear{2010Li}).
Note that this result is not asymptotic and holds for any $p$ and
$q<p$. From this theorem, under the regularity assumption~(A1)
and the following further assumptions on $p$ and the tuning parameters
$\lambda_1, \lambda_2$:
\begin{longlist}[(A.3)]
\item[(A3)] $\lim_{n \rightarrow\infty} \frac{p}{n}=0$,

\item[(A4)] $\lim_{n \rightarrow\infty}\frac{\lambda_1\sqrt
{p}}{n}=0$,

\item[(A5)] $\lim_{n \rightarrow\infty}\frac{\lambda_2}{n}=0$ and
$\lim_{n \rightarrow\infty}\frac{\lambda_2\|\beta
_{(1)}\|_2}{n}=0$,
\end{longlist}
we have
\[
\|\hat{\beta}(\lambda_1,\lambda_2)-\beta\|_2^2 \stackrel
{P}{\longrightarrow} 0,
\]
which implies that the Grace of $\beta$ is $L_2$ consistent. This
result implies that the Grace procedure chooses the important
variables with high probability and that falsely chosen variables by
Grace have very small coefficients. The $L_2$ consistency result
suggests that we may use some hard-thresholding procedure to further
eliminate the variables with very small Grace coefficients.
Alternatively, an interesting randomized selection procedure
proposed by Bickel, Ritov and Tsybakov (\citeyear{2008Bickel}) can be used to further
eliminate the variables with small estimated Grace coefficients.
Note that under the classical setting where $p$, $q$ and $\beta_i$
are all fixed as $n\rightarrow\infty$, the assumptions (A3)--(A5) hold
when $\lambda_i/n \rightarrow\infty, i=1,2.$

\subsection{Model selection consistency when $p$ is fixed}

We next establish the results on model selection consistency for the
standard case where $p$ and $q$ are fixed when $n \rightarrow
\infty$. Following Zhao and Yu (\citeyear{2006Zhao}), we define the Grace of
$\beta$ to be sign consistent if there exists $\lambda_1$ and
$\lambda_2$ as functions of $n$ such that
\begin{eqnarray*}
\lim_{n\rightarrow
\infty}\operatorname{Pr}\bigl(\operatorname{sign}(\hat{\beta}(\lambda_1,\lambda
_2))=\operatorname{sign}(\beta)\bigr)=1.
\end{eqnarray*}
To establish the sign consistency of the Grace, we first provide the
following graph-constrained irrepresentable condition (GC-IC): there
exists $\eta>0$ and $\lambda_1>0$, \mbox{$\lambda_2>0$}, such that
\begin{eqnarray}\label{GC-IC}
\quad &&\biggl|\biggl(\mathrm{C}_{21}+\frac{\lambda_2}{n}\mathrm
{L}_{21}
\biggr)\biggl(\mathrm{C}_{11}+\frac{\lambda_2}{n}\mathrm{L}_{11}\biggr)^{-1}
\biggl[\operatorname{sign}\bigl(\beta_{(1)}\bigr)+\frac{2\lambda_2}{\lambda
_1}\mathrm{L}
_{11}\beta_{(1)}\biggr]\nonumber
\\
&&{}\qquad\quad\hspace*{178pt} -\frac{2\lambda_2}{\lambda_1}\mathrm{L}_{21}\beta_{(1)}
\biggr|
\\
&&\qquad \le \mathbf{1}-\eta,\nonumber
\end{eqnarray}
where $\mathbf{1}$ is a vector of 1s with length $p-q$ and the
inequality holds element-wise. Further, we assume that
$\mathrm{C}\rightarrow\mathrm{C}_0$, where $\mathrm{C}_0$ is
positive definite. The
GC-IC is a consequence of the Karush--Kuhn--Tucker (KKT) conditions
for the following constrained optimization problem that corresponds
to the penalized optimization problem of equation (\ref{Network}):
\[
\hat{\beta}(\lambda,\alpha)=\operatorname{argmin}\limits_{\beta}
\biggl\{\frac{1 }{2n}\|\mathrm{y}-\mathrm{X}\beta\|_2^2\dvtx P_{\alpha
}(\beta)\le
\lambda\biggr\}.
\]

\begin{theorem}\label{thm3.2}
For fixed $p,q$ and $\beta$, if $\mathrm{C}\rightarrow\mathrm
{C}_0$, where
$\mathrm{C}_0$ is positive definite and condition \textup{(A.2)} holds, the
graph-constrained estimate is sign consistent if and only if GC-IC
\textup{(\ref{GC-IC})} holds for $\lambda_1, \lambda_2$ that satisfy
$\lambda_1/\sqrt{n} \rightarrow\infty$ and $\lambda_i/n\rightarrow
0$, for~$i=1,2$.
\end{theorem}

This theorem is a special case of Theorem~\ref{thm3.3} and its proof is
similar to that of Zhao and Yu (\citeyear{2006Zhao}) for Lasso estimates. We
therefore omit its proof in this paper. Note that the required
conditions on the sparsity tuning parameter $\lambda_1$ are the
same as those for the Lasso [Zhao and Yu (\citeyear{2006Zhao})], for example,
$\lambda_1=\sqrt{n}\log n$ is a suitable choice. This theorem
indicates that under some restrictive conditions of the design
matrix and the Laplacian matrix of the network, the sign consistency
property holds for the graph-constrained regularization. To gain
further insight into GC-IC, consider the special cases when
$\lambda_2$ is preselected and fixed and when $\lambda_1$ goes to
infinity, the GC-IC reverses back to the irrepresentable condition
for the Lasso given in Zhao and Yu (\citeyear{2006Zhao}) and the graph-constrained
penalty function $\lambda_1\|\beta\|_1+\lambda_2\beta^T \mathrm{L}
\beta=\lambda_1(\|\beta\|_1+\frac{\lambda_2 }{\lambda_1}\beta^T
\mathrm{L}
\beta)$ is reduced to the Lasso penalty.

\subsection{Model selection consistency when $p$ diverges}

We now consider the model selection consistency of the
graph-constrained regularization procedure under the settings when
the number of covariates $p$ also goes to infinity as $n
\rightarrow\infty$, in which case, the assumptions and the
regularity conditions for Theorems~\ref{thm3.1} and~\ref{thm3.2} become inappropriate
as $\mathrm{C}$ does not converge and $\beta$ may change as $n$
grows. The
following theorem shows
that for the general
scalings when $p,q$ and~$n$ all go to infinity, under some
additional conditions between $p,q$ and $n$, GC-IC also guarantees
that the Grace is sign consistent in selecting the true model.

\begin{theorem}\label{thm3.3}
Suppose each column of $\mathrm{X}$ is normalized to the $L_2$-norm of
$n$ and
GC-IC (\ref{GC-IC}) holds. Define
$\rho:=\min|(\mathrm{C}_{11}+\frac{\lambda_2}{n}\mathrm{L}
_{11})^{-1}(\mathrm{C}_{11}\beta_{(1)})|$
and $\mathrm{C}_{\min}=\Lambda_{\min}(\mathrm
{C}_{11})$, where
$\Lambda_{\min}(\cdot)$ denotes the minimal eigenvalue. Let
$W_{\max}=\max_{u,v}\{w(u,v)\}$. If $\lambda_1$ and
$\lambda_2$ are chosen such that:
\begin{longlist}[(a)]
\item[(a)] If $\mathrm{L}_{12}=0$,
\[
\frac{\lambda_1^2}{n\log(p-q)} \rightarrow\infty,
\]
or if $\mathrm{L}_{12}\ne0$,
\[
\frac{\lambda_1^2}{\log(p-q)(n+{\lambda_2^2 W_{\max}^2/(n\mathrm{C}_{\min})})}
\rightarrow\infty.
\]

\item[(b)] If $\frac{1}{\rho}\{\sqrt{\frac{\log
q}{n\mathrm{C}_{\min}}}+
\frac{\lambda_1}{n}\|(\mathrm
{C}_{11}+\frac
{\lambda_2}{n}\mathrm{L}_{11})^{-1}\operatorname{sign}
(\beta_{(1)})\|_{\infty}\}\rightarrow
0$,
\end{longlist}
then the Grace $\hat{\beta}(\lambda_1,\lambda_2)$ is sign consistent
as $n\rightarrow\infty$.
\end{theorem}

A proof analogous to Jia and Yu (\citeyear{2008Jia}) can be found in Li and Li
(\citeyear{2010Li}). Theorem~\ref{thm3.3} gives a general sign consistency result for
the Grace for general scalings of $p,q$ and $n$. If
$\mathrm{C}_{\min}\ge\alpha$ for some $\alpha>0$ and $\rho
\le
\rho_0$ for some $\rho_0>0$, it is easy to check that the conditions
$\log q/n \rightarrow\infty$ and $\lambda_1\sqrt{q}/n \rightarrow
0$ guarantee that condition (b) in Theorem~\ref{thm3.3} holds. In the
settings when $p$ and $q$ are fixed, if $\mathrm{C}_{11}$ converges to a
nonnegative definite matrix $\mathrm{C}_{110}$, $\rho$ converges to a
nonnegative number. In addition, it is easy to verify that the
conditions in Theorem~\ref{thm3.2} guarantee that the conditions (a) and (b)
in Theorem~\ref{thm3.3} hold.

\section{Monte Carlo simulations}\label{simu}
We conducted Monte Carlo simulations to evaluate the proposed Grace
and aGrace procedures and to compare the performance of this
procedure with Lasso and Enet in terms of prediction errors and
identification of relevant
variables.
We simulated the graph to mimic gene regulation modules. We used
genes and variables interchangeably in this section. We assumed that
the graph consisted of 200 unconnected regulatory modules with 200
transcription factors (TFs) and each regulated 10 different genes
for a total of 2200
variables. 
Among these modules and genes, we further assumed that
four TFs and their 10 regulated genes (for a total of 44 variables)
were associated with the response based on the following model:
%
\begin{eqnarray}\label{simu.model}
Y=\sum_{u=1}^{44} \beta_u X_u +\epsilon.
\end{eqnarray}

We considered two different models. For the first model, we assumed
that the coefficients in model (\ref{simu.model}) were specified as
\begin{eqnarray*}
\beta&=&\biggl(2, \underbrace{\frac{2}{\sqrt{10}},\ldots,\frac{2}{
\sqrt{10}}}_{10},-2, \underbrace{\frac{-2}{\sqrt{10}},\ldots
,\frac{-2}{
\sqrt{10}}}_{10},
4,
\\
&&{}\; \underbrace{\frac{4}{\sqrt{10}},\ldots,\frac{4}{
\sqrt{10}}}_{10}, -4, \underbrace{\frac{-4}{\sqrt{10}},\ldots,\frac{-4}{
\sqrt{10}}}_{10},0,\ldots,0\biggr),
\end{eqnarray*}
and the $\epsilon$ was
random mean-zero normal error with variance $\sigma^2=\sum_u
\beta_u^2/4$. For each TF, the $\mathrm{X}$ value was simulated from a
$N(0,1)$ distribution, and conditional on the value of the TF, we
simulated the expression levels of the genes that they regulated
from a conditional normal distribution with correlations of 0.2, 0.5
and 0.9, respectively. We therefore had a total of 2200 variables
and 44 of them were relevant. For the second model, we considered
the case when the regulated genes had different signs in regression
coefficients, where the regression coefficients in model
(\ref{simu.model}) have the same absolute values as in Model 1, but
for each simulated module, three out of the 10 genes regulated by
the TF had different signs from the other 7 genes.
The $\mathrm{X}$ values were generated in the same way as in previous
simulations. In this model, genes that are regulated by the same
transcription factor are assumed to have different regression
coefficients.

For each model, our simulated data consisted of a training set, an
independent validation set and an independent test set with a sample
size of 200 for all three data sets. Models were fitted on training
data only, and the validation data were used to select the tuning
parameters. We computed the prediction mean-squared errors on the
test data set. For each model, we repeated the simulations 100 times.
Table \ref{Box} shows the prediction mean-square errors for several
different procedures. For Model 1 when the neighboring genes have
the same signs in regression coefficients, we observed that the
Grace gave~the smallest prediction errors for all four models with
different correlations among the predictors. Both Grace and Enet
performed better than Lasso in prediction. When the correlation is
very high, the prediction errors of these procedures were
comparable, however, Grace still gave~the smallest prediction error
among the~procedures compared. When the signs of the~regression
coefficients were the same, aGrace was reduced to Grace and gave
the same prediction results.
For Model 2 when the neighboring variables have different signs of
coefficients, aGrace adjusting for the signs of the regression
coefficients gave the smallest prediction errors, further indicating
the importance
of adjusting for the signs in the regularization. In
general, Grace gave similar prediction results as the Enet, except
when the correlation between the transcription
factors and their regulated genes was very high, in which case Enet
resulted in a slightly smaller prediction error.

\begin{table}
\caption{Comparison of prediction mean-square errors
(SE) using
Grace, aGrace, Enet and Lasso for three different correlation
structures of 0.2, 0.5 and 0.9 between the transcription factors and
their regulated genes for each of the two models considered. The
results are based on 100 replications}\label{Box}
\begin{tabular*}{\tablewidth}{@{\extracolsep{4in minus 4in}}lcccccc@{}}
\hline
 & \multicolumn{3}{c}{\textbf{Model 1 (Cor)}}& \multicolumn{3}{c@{}}{\textbf{Model 2 (Cor)}}\\[-6pt]
  & \multicolumn{3}{c}{\hrulefill}& \multicolumn{3}{c@{}}{\hrulefill}\\
\textbf{Method} & \textbf{0.2}& \textbf{0.5} & \textbf{0.9} & \textbf{0.2}& \textbf{0.5}& \textbf{0.9}\\
\hline
Grace & 24.93 & 23.22 & 22.56  &53.08 & 42.07 & 28.20 \\
& (2.97) & (2.41)& (2.20) & (6.45) & (5.03)& (2.87) \\
aGrace & 24.93 & 23.22 & 22.56 & 27.70 & 26.23 & 25.55 \\
& (2.97) & (2.41)& (2.20)& (3.66) & (3.03) & (2.76) \\
Enet& 51.33 & 37.37 & 25.82  &56.18 & 45.65 & 27.33 \\
& (6.65) & (4.69) & (2.67) & (7.22)& (5.81) & (2.72) \\
Lasso& 53.41 & 40.30 & 27.82 & 57.62& 47.65 & 29.23 \\
& (6.68)& (4.94) & (2.98) & (7.01) & (5.53) & (2.78) \\
\hline
\end{tabular*}
\end{table}

\begin{figure}

\includegraphics{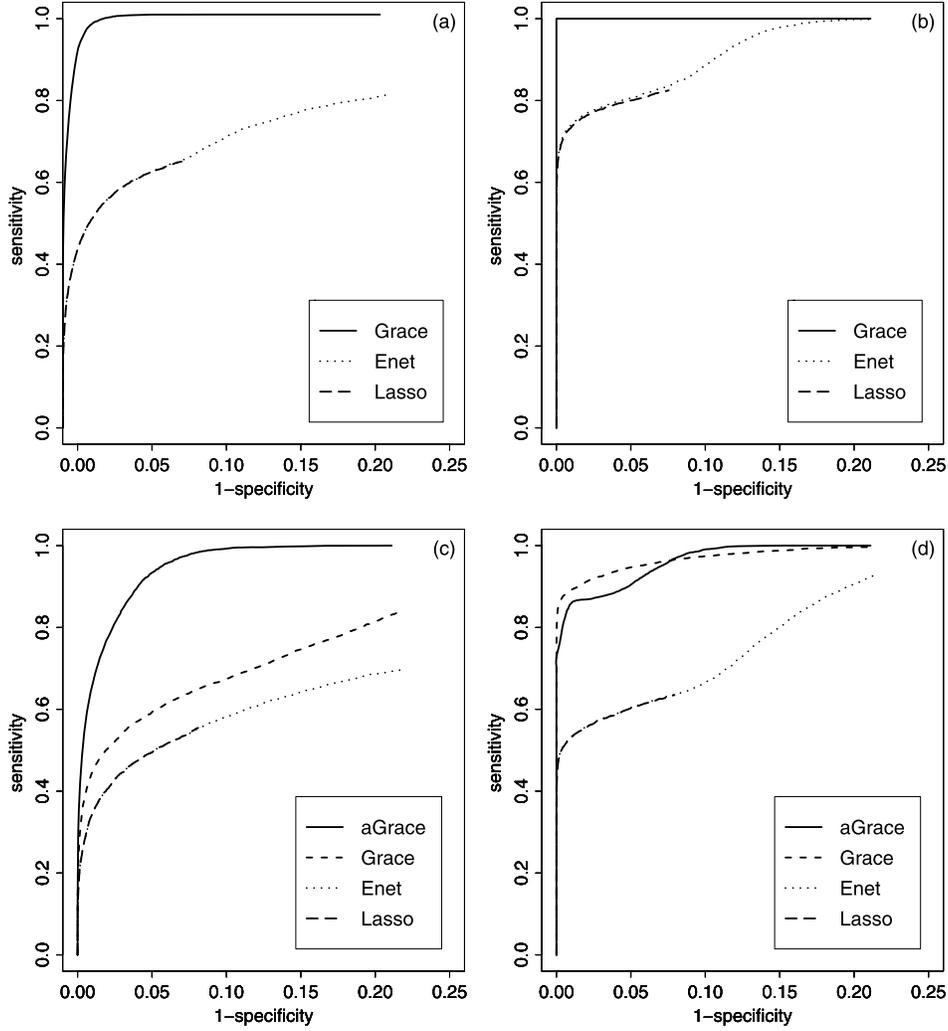}

\caption{Comparison of ROCs for Grace, aGrace, Enet and Lasso for
Model 1 [plots \textup{(a)} and \textup{(b)}] and Model 2 [plots \textup{(c)} and \textup{(d)}] and for
correlations of 0.2 [plots \textup{(a)} and \textup{(c)}] and 0.9 [plots \textup{(b)} and~\textup{(d)}].
The ROCs were calculated as a function
of the sparsity parameter $\lambda_1$. For Grace, aGrace~and Enet, the
tuning parameter $\lambda_2$
was selected based on 5-fold CV.}\label{ROC}
\end{figure}

To compare the performance on variable selection, Figure \ref{ROC}
shows the receiver operating characteristic (ROC) curves of several
different procedures in selecting the relevant variables for the
models with correlation of 0.2 and 0.9 between the TF and their
regulated genes. For Grace, aGrace and Enet, the ROC curves were
obtained as a function of the sparsity parameter $\lambda_1$ with
tuning parameter $\lambda_2$ selected based on 5-fold
cross-validation among the values of 0.1, 1, 10, 100 and 1000.
For Model 1 when the neighboring genes have the same signs in
regression coefficients [Figure \ref{ROC} plots (a) and (b)], Grace
gave much larger areas under the ROC cruves than Enet and Lasso,
indicating better performance in variable selection for Grace. In
addition, five-fold cross-validation always chose the largest
$\lambda_2=1000$ for Grace and $\lambda_2=0.1$ for Enet in all 100
replications. For Model 2 when the neighboring variables have
different signs of coefficients, aGrace adjusting for the signs of
the regression coefficients performed better than the other three
procedures compared, resulting in larger areas under the curves, and
Grace still performed better than Lasso and Enet on variable
selections in both low and high correlation scenarios. When the
correlation among the relevant variables is low, the 5-fold CV
selected $\lambda_2=1000$ for aGrace and $\lambda_2=0.1$ for Enet in
all 100 replications and selected $\lambda_2=100$ for Grace in most
of the replications. When the correlation among the relevant
variables was high, the 5-fold CV selected $\lambda_2=1000$ for
aGrace and $\lambda_2=100$ for Grace in most of the 100
replications and selected $\lambda_2=0.1$ for Enet in all the
replications.

\section{Application to network-based analysis of gene expression data}\label{realdata}

To demonstrate the proposed method, we consider the problem of
identifying age-dependent molecular modules based on the gene
expression data measured in human brains of individuals of different
ages published in Lu et~al. (\citeyear{2004Lu}). In this study the gene
expression levels in the postmortem human frontal cortex were
measured using the Affymetrix arrays for 30 individuals ranging from
26 to 106 years of age.
To identify
the aging-regulated genes, Lu et~al. (\citeyear{2004Lu}) performed simple
linear regression analysis for each gene with age as a covariate. We
analyzed this data set by combining the KEGG regulatory network
information with the gene expression data [Kanehisa and Goto (\citeyear{2002Kanehisa})].
In particular, we limited
our analysis to the genes that can be mapped to the KEGG regulatory
work and focused on
the problem of identifying the subnetworks of the KEGG regulatory
network that are associated with
human brain aging. By merging the gene expression data with the KEGG
regulatory pathways,
the final KEGG network includes 1305 genes and 5288 edges.

\begin{table}[b]
\caption{Results of analysis of brain aging gene expression data
by four different procedures, including the number of genes selected
(No. genes), the number of linked KEGG edges (No. edges), the five-fold
cross-validation error (CV error) and the values of the tuning
parameters  selected($\lambda_2$ for Grace, aGrace and
Enet and $s_1=\sum_v |\beta_v|$)}\label{tab2}
\begin{tabular*}{\tablewidth}{@{\extracolsep{4in minus 4in}}lcccc@{}}
\hline
& \textbf{No. genes} & \textbf{No. edges} &\textbf{CV error} & \textbf{Tuning parameters} \\
\hline
Grace & 45 & \phantom{0}9 & 0.079 & $\lambda_2=0.1$, $s_1= 4.72$\phantom{1}\\
aGrace &73 & 28 & 0.077 & $\lambda_2=0.01$, $s_1=6.97$\\
Enet & 41& 10& 0.077 & $\lambda_2=1.0$, $s_1=5.64$\phantom{1}\\
Lasso & 18 & \phantom{0}0 & 0.098 & $s_1=5.65$\phantom{ $s_1=5.64$}\\
\hline
\end{tabular*}
\end{table}

\begin{figure}

\includegraphics{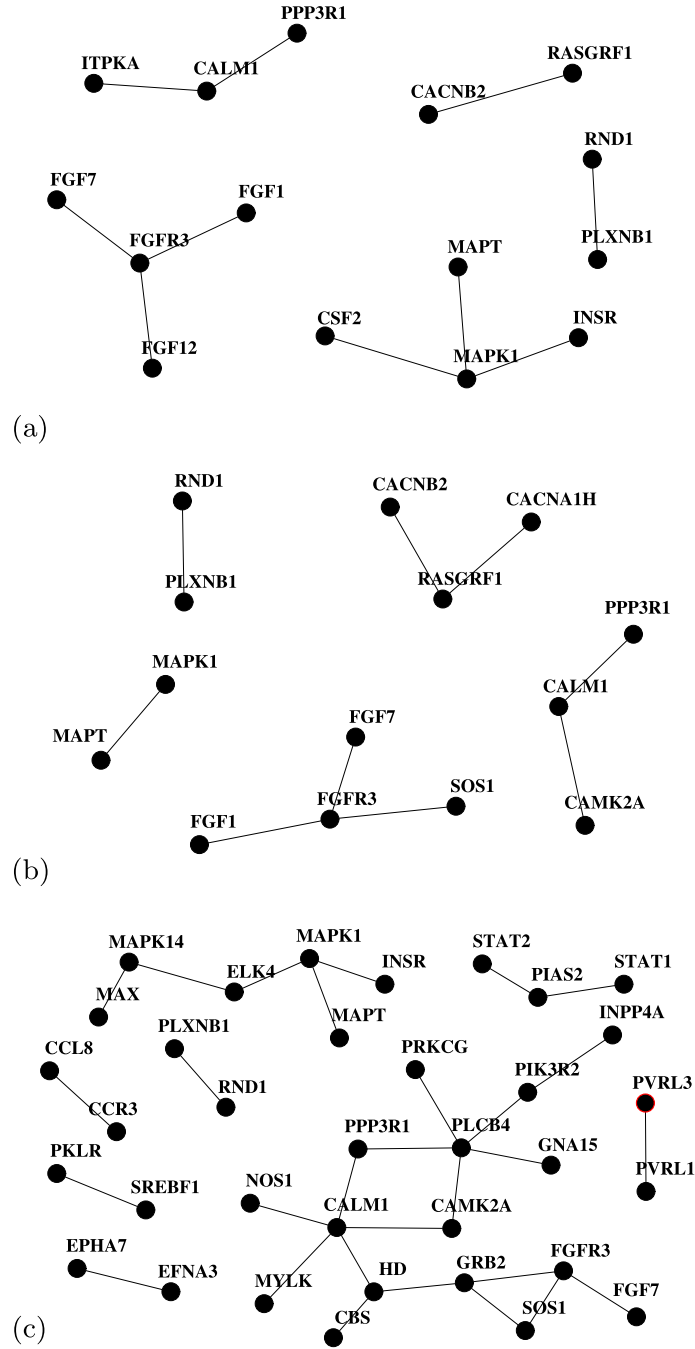}

\caption{Subnetworks identified by
\textup{(a)} Elastic net (Enet), \textup{(b)} Grace and \textup{(c)} aGrace for brain aging
gene expression data (only those genes that are linked on the KEGG
network are plotted).} \label{brain.fig}
\end{figure}

We treated the logarithm of the individual age as the response variable
and the expression levels (after $\log_{10}$ transformation) of 1305
genes on the KEGG network as the explanatory variables in our analysis.
Table~\ref{tab2} shows the results of several different procedures where the
tuning parameters were selected
by five-fold cross-validations. Overall, we observed that the Lasso
selected the fewest number of genes with relatively large
cross-validation errors and Grace and Enet selected roughly the same
number of genes with similar CV errors. However, the adaptive Grace
resulted in more identified genes with similar CV errors than the
other two procedures. Figure~\ref{brain.fig} shows the subnetworks
identified by four different estimation procedures. As a comparison,
we also included the genes selected by Lasso, although it did not
select any linked pairs of genes on the KEGG network. It is
interesting to note that as we impose more constraints on the
regression coefficients, more linked genes are identified. Both
Enet and Grace identified some common subnetworks that were
associated with brain aging. These included fibroblast growth
factors (FGF) and its receptors. It has been demonstrated that FGFs
are associated with many developmental processes including neural
induction [Bottcher and Niehrs (\citeyear{2005Bottcher})]
and are involved in multiple functions including cell proliferation,
differentiation, survival and aging [Yeoh and
de Haan (\citeyear{2007Yeoh})]. It is also interesting to observe
that mitogen-activated protein kinase (MAPK) (MAPK1 and MAPK9) and
the specific MAPK kinase (MAP2K) were also identified by Enet and
Grace. The MAPKs play important roles in induction of apoptosis
[Hayesmoore et~al. (\citeyear{2010Hayesmoore})]. Other interesting genes include
RAS protein-specific guanine nucleotide-releasing factor 1 (RASGRF1),
the functionality of which is highly significant in various contexts of
the central nervous system.
In the hippocampus, RASGRF2 has been shown to interact with the NR2A
subunits of NMDARs, triggering Ras-ERK activation and induction of
long-term potentiation, a form of neuronal plasticity that contributes
to memory storage in the brain [Tian et~al. (\citeyear{2004Tian}); Lu
et~al. (\citeyear{2004Lu})]. Finally, the insulin receptor gene (INSR) is
also identified.
INSR binds insulin (INS) and regulates energy metabolism. Evidence from
model organisms, including results from fruit flies [Tatar et
al. (\citeyear{2001Tatar})]
and roundworms [Kimura et~al. (\citeyear{1997Kimura})], relates INSR
homologues to
aging, most likely as part of the GH1/IGF1 axis. These results
indicated that our method
can indeed recover some biologically interesting molecular modules or
KEGG subnetworks that
are related to brain aging in humans.

It is important to point out that the adaptive Grace identified
several small sets of connected genes that were missed by Enet or
the standard Grace. One of the subnetworks included EPHRIN and Eph
receptor, both of which were found to be related to neural
development and entohino-hippocampal axon targeting [Flanagan and
Vaderhaeghen (\citeyear{1998Flanagan}); Stein  et al. (\citeyear{1999Stein})]. Another subnetwork
was part of the Jak-State signaling, which is important in both
mature and aging brains [De-Frajaa et~al. (\citeyear{2000De})]. Aging was
also found to be associated with increased human T cell CC chemokine
receptor gene expression [Yung et~al. (\citeyear{2003Yung})]. Other
interesting subnetworks included PVRL3--PVRL1 that are associated
with cell adhesion.




\section{Discussion}
We have introduced and studied the theoretical properties of a
graph-constrained regularized estimation procedure for linear
regressions in order to incorporate information coded in graphs.
Such a regularization procedure can also be regarded as a penalized
least squared estimation where the penalty is defined as a
combination of the $L_1$ and $L_2$ penalties on degree-scaled
differences of coefficients between variables linked on the graphs.
This penalty function induces both sparsity and smoothness with
respect to the graph structure of the regression coefficients.
Simulation studies indicated that when the coefficients are similar
for variables that are neighbors on the graph, the proposed
procedure has better prediction and identification performance than
other commonly used regularization procedures such as Lasso and
elastic net regressions. Such improvement results from effectively
utilizing the neighboring information in estimating the regression
coefficients. If the smoothness assumption on the coefficients does
not hold, we expect that the cross-validation selects a very small
value of $\lambda_2$ and, therefore, the proposed procedure would
perform similarly as the Lasso.
In analysis of the brain aging gene expression data, different
from Lasso, the new procedure tends to identify sets of linked genes
on the networks, which often leads to better biological
interpretation of the genes identified. Although the methods
presented are largely motivated by applications in genomic data,
they can be applied to other settings when the covariates are nodes
on general graphs, such as in image analysis.

Although the methods presented in this paper were developed mainly
for linear models, similar methods can be developed for the
generalized linear models and the censored survival data regression
models, where we can use the negative of the logarithm of the
likelihood or partial likelihood as the loss function. Similar to
the techniques presented in Friedman et~al. (\citeyear{2007Friedman}) and Wu and
Lange (\citeyear{2008Wu}), we can use the coordinate descent procedure together
with the iterative reweighted least square to obtain the solution
path. Such models have great applications in genomic data analysis
in identifying the genes or subnetworks that are associated with
binary or censored survival data outcomes. Other extensions include
replacing the $L_1$ part of the Grace penalty with other sparse
penalty functions such as SCAD or bridge penalty
[Huang et~al. (\citeyear{2008Huang})]. Important future research also includes how to handle
covariates that are linked on directed graphs. Finally, to
incorporate the fact that the linked nodes might be negatively
correlated and the corresponding regression coefficients may have
different signs, we introduced an adaptive sign-adjusted
graph-constrained regularization procedure and showed that such a
procedure can perform better than the original graph-constrained
regularization. The theoretical property of such an adaptive
procedure is unknown and is an area for future research.

\section*{Acknowledgments}
 We thank the two reviewers for their comments
that have greatly improved the presentation of this paper.

\begin{supplement}[id-suppA]
\stitle{Proofs of Theorem \ref{thm3.1} and Theorem \ref{thm3.3}}
\slink[doi]{10.1214/10-AOAS332SUPP}
\slink[url]{http://lib.stat.cmu.edu/aoas/332/supplement.pdf}
\sdatatype{.pdf}
\sdescription{We present the details of the
proofs  of Theorem \ref{thm3.1} and Theorem \ref{thm3.3} in the Supplemental Materials.}
\end{supplement}

\printaddresses


\begin{thebibliography}{99}



\bibitem[\protect\citeauthoryear{}{2008}]{2008Bickel}
\textsc{Bickel, P. L., Ritov, Y.} and \textsc{Tsybakov, A. B.} (2008).
Hierarchical selection of variables in sparse
high-dimensional regression. {Technical report, Dept. Statistics, Univ. California, Berkeley.}

\bibitem[\protect\citeauthoryear{}{2005}]{2005Bottcher}
\textsc{Bottcher, R. T.} and \textsc{Niehrs, C.} (2005). Fibroblast
growth factor
signaling during early vertebrate development. \textit{Endocrine
Reviews} \textbf{26} 63--77.

\bibitem[\protect\citeauthoryear{}{1997}]{1997Chung}
\textsc{Chung, F.} (1997). \textit{Spectral Graph Theory}.
 \textit{CBMS Reginal Conferences Series} \textbf{92}. Amer. Math. Soc.,
Providence, RI.
\MR{1421568}


\bibitem[\protect\citeauthoryear{}{2000}]{2000De}
\textsc{De-Fraja, C., Conti, L., Govoni, S., Battaini, F.} and \textsc
{Cattaneo, E.} (2000).
STAT signalling in the mature and aging brain.
\textit{International Journal of Developmental Neuroscience} \textbf
{18} 439--446.

\bibitem[\protect\citeauthoryear{}{1994}]{1994Donoho}
\textsc{Donoho, D.} and \textsc{Johnstone, I.} (1994). Ideal spatial
adaptation via wavelet shrinkage.
\textit{Biometrika} \textbf{81} 425--455.
\MR{1311089}

\bibitem[\protect\citeauthoryear{}{2004}]{2004Efron}
\textsc{Efron, B., Hastie, T., Johnstone, I.} and \textsc{Tibshirani, R.}
(2004). Least angle regression. \textit{Ann. Statist.} \textbf{32}
407--499.
\MR{2060166}

\bibitem[\protect\citeauthoryear{}{2001}]{2001Fan}
\textsc{Fan, J.} and \textsc{Li, R.} (2001). Variable selection via nonconcave
penalized likelihood and its oracle properties. \textit{J.
Amer. Statist. Assoc.} \textbf{96} 1348--1360.
\MR{1946581}

\bibitem[\protect\citeauthoryear{}{2004}]{2004Fan}
\textsc{Fan, J.} and \textsc{Peng, H.} (2004). Nonconcave penalized
likelihood with a
diverging number of parameters. \textit{Ann. Statist.} \textbf{32} 928--961.
\MR{2065194}

\bibitem[\protect\citeauthoryear{}{1998}]{1998Flanagan}
\textsc{Flanagan, J. G.} and \textsc{Vanderhaeghen, P.} (1998). The
ephrins and Eph
receptors in neural development. \textit{Annual Review Neuroscience}
\textbf{21} 309--345.

\bibitem[\protect\citeauthoryear{}{2007}]{2007Friedman}
\textsc{Friedman, J., Hastie, T., Hoefling, H.} and \textsc
{Tibshirani, R.}
(2007). Pathwise coordinate optimization. \textit{Ann. Appl.
Statist.} \textbf{1} 302--332.
\MR{2415737}


\bibitem[\protect\citeauthoryear{}{2009}]{2010Hayesmoore}
\textsc{Hayesmoore, J. B., Bray, N. J., Cross, W. C., Owen, M. J.,
O'Donovan, M. C.} and
\textsc{Morris, H. R.} (2009).
The effect of age and the H1c MAPT haplotype on MAPT expression in
human brain.
\textit{Neurobiol. Aging} \textbf{30} 1652--1656.



\bibitem[\protect\citeauthoryear{}{2008}]{2008Huang}
Huang, J., Horowitz, J. L. and Ma, S. (2008). Asymptotic properties of
bridge estimators in sparse
high-dimensional regression models. \textit{Ann. Statist.}
\textbf{36} 587--613.

\bibitem[\protect\citeauthoryear{}{2007}]{2007Huang}
\textsc{Huang, J.} and \textsc{Xie, H.} (2007). Asymptotic oracle
properties of SCAD-penalized least
squares estimators. In \textit{Asymptotics: Particles, Processes and
Inverse Problems. IMS Lecture Notes  Monogr. Ser.} \textbf{55} 149--166.
IMS, Beachwood, OH.
\MR{2459937}


\bibitem[\protect\citeauthoryear{}{2008}]{2008Jia}
\textsc{Jia, J.} and \textsc{Yu, B.} (2008). On model selection
consistency of
elastic net when $p \gg n$. {Technical Report 756, Dept. Statistics, Univ. California,
Berkeley.}

\bibitem[\protect\citeauthoryear{}{2002}]{2002Kanehisa}
\textsc{Kanehisa, M.} and \textsc{Goto, S.} (2002). KEGG: Kyoto
encyclopedia of
genes and genomes. \textit{Nucleic Acids Research} \textbf{28} 27--30.

\bibitem[\protect\citeauthoryear{}{1997}]{1997Kimura}
\textsc{Kimura, K. D., Tissenbaum, H. A., Liu, Y.} and \textsc
{Ruvkun, G.} (1997).
daf-2, an insulin receptor-like gene that regulates longevity and
diapause in Caenorhabditis elegans.
\textit{Science} \textbf{277} 942--946.




\bibitem[\protect\citeauthoryear{}{2008}]{2008Li}
\textsc{Li, C.} and \textsc{Li, H.} (2008). Network-constrained regularization
and variable selection
for
analysis of genomic data. \textit{Bioinformatics} \textbf{24} 1175--1182.



\bibitem[\protect\citeauthoryear{}{2010}]{2010Li}
\textsc{Li, C.} and \textsc{Li, H.} (2010). Supplement to ``Variable
selection and regression analysis for
graph-structured covariates with an application to genomics'' DOI:
\href{http://dx.doi.org/10.1214/10-AOAS332SUPP}{10.1214/10-AOAS332SUPP}.


\bibitem[\protect\citeauthoryear{}{2004}]{2004Lu}
\textsc{Lu, T., Pan, Y., Kao, S.-Y., Li, C., Kohane, I.,
Chan, J.} and \textsc{Yankner, B. A.} (2004). Gene regulation and DNA damage
in the aging human brain. \textit{Nature} \textbf{429} 883--891.



\bibitem[\protect\citeauthoryear{}{2008}]{2008Nardi}
\textsc{Nardi, Y.} and \textsc{Rinado, A.} (2008). On the asymptotic
properties of the group lasso estimator for linear models.
\textit{Electron. J. Statist.} \textbf{2} 605--633.
\MR{2426104}

\bibitem[\protect\citeauthoryear{}{1984}]{1984Portnoy}
\textsc{Portnoy, S.} (1984). Asymptotic behavior of M-estimators of $p$
regression parameters when $p/n$ is large. I. Consistency.
 \textit{Ann. Statist.} \textbf{12} 1298--1309.
\MR{0760690}

\bibitem[\protect\citeauthoryear{}{1999}]{1999Stein}
\textsc{Stein, E., Savaskan, N. E., Ninnemann, O., Nitsch, R., Zhou,
R.} and
\textsc{Skutella, T.} (1999). A role for the Eph ligand ephrin-A3 in
entorhino-hippocampal axon targeting. \textit{Journal of Neuroscience}
\textbf{19} 8885--8893.

\bibitem[\protect\citeauthoryear{}{2001}]{2001Tatar}
\textsc{Tatar, M., Kopelman, A., Epstein, D., Tu, M. P., Yin, C. M.}
and \textsc{Garofalo, R. S.} (2001).
A mutant Drosophila insulin receptor homolog that extends life-span and
impairs neuroendocrine function.
\textit{Science} \textbf{292} 107--110.

\bibitem[\protect\citeauthoryear{}{2004}]{2004Tian}
\textsc{Tian, X., Gotoh, T., Tsuji, K., Lo, E. H., Huang, S.} and
\textsc{Feig, L. A.} (2004).
Developmentally regulated role for Ras-GRFs in coupling NMDA glutamate
receptors to Ras, Erk and CREB. \textit{EMBO J.}
\textbf{23} 1567--1575.

\bibitem[\protect\citeauthoryear{}{1996}]{1996Tibshirani}
\textsc{Tibshirani, R. J.} (1996). Regression shrinkage and selection via
the lasso. \textit{J. Roy.
Statist. Soc. Ser. B} \textbf{58} 267--288.
\MR{1379242}

\bibitem[\protect\citeauthoryear{}{2005}]{2005Tibshirani}
\textsc{Tibshirani, R., Saunders, M., Rosset, S., Zhu, J.} and \textsc
{Knight, K.} (2005). Sparsity and smoothness via
the fused lasso. \textit{J. Roy. Statist. Soc. Ser. B}
\textbf{67} 91--108.
\MR{2136641}


\bibitem[\protect\citeauthoryear{}{2008}]{2008Wu}
\textsc{Wu, T. T.} and \textsc{Lange, K.} (2008). Coordinate descent
algorithms for lasso penalized regression.
\textit{Ann. Appl. Statist.} \textbf{2} 224--244.
\MR{2415601}

\bibitem[\protect\citeauthoryear{}{2007}]{2007Yeoh}
\textsc{Yeoh, J. S.} and \textsc{de Haan, G.} (2007).
Fibroblast growth factors as regulators of stem cell self-renewal and aging.
\textit{ Mechanisms of Ageing and Development} \textbf{128} 17--24.

\bibitem[\protect\citeauthoryear{}{2006}]{2006Yuan}
\textsc{Yuan, M.} and \textsc{Lin, Y.} (2006). Model selection and
estimation in
regression with grouped variables. \textit{J. Roy.
Statist. Soc. Ser. B} \textbf{68} 49--67.
\MR{2212574}

\bibitem[\protect\citeauthoryear{}{2003}]{2003Yung}
\textsc{Yung, R. L.} and \textsc{Mo, R.} (2003).
Aging is associated with increased human T cell CC chemokine
receptor gene expression. \textit{Journal of Interferon \& Cytokine
Research} \textbf{23} 575--582.

\bibitem[\protect\citeauthoryear{}{2006}]{2006Zhang}
\textsc{Zhang, C.} and \textsc{Huang, J.} (2006). The sparsity and
bias of the Lasso selection
in high-dimensional linear regression. \textit{Ann. Statist.}
\textbf{36} 1567--1594.
\MR{2435448}

\bibitem[\protect\citeauthoryear{}{2006}]{2006Zhao}
\textsc{Zhao, P.} and \textsc{Yu, B.} (2006). On model selection
consistency of Lasso. \textit{J.
Mach. Learn. Res.} \textbf{7} 2541--2567.
\MR{2274449}

\bibitem[\protect\citeauthoryear{}{2004}]{2004Zhou}
\textsc{Zhou, D., Bousquet, O., Lal, T., Weston, J.} and \textsc
{Scholkopf, B.} (2004).
Learning with local and global consistency. In \textit{NIPS}
\textbf{16} 321--328.  MIT Press, Cambridge, MA.

\bibitem[\protect\citeauthoryear{}{2005}]{2005Zhu}
\textsc{Zhu, X.} (2005). Semi-supervised learning literature survey.
{Technical Report 1530, Computer
Sciences, University of Wisconsin--Madison}.

\bibitem[\protect\citeauthoryear{}{2006}]{2006Zou}
\textsc{Zou, H.} (2006).
The adaptive lasso and its oracle properties.
\textit{J. Amer. Statist. Assoc.} \textbf{101} 1418--1429.
\MR{2279469}

\bibitem[\protect\citeauthoryear{}{2005}]{2005Zou}
\textsc{Zou, H.} and \textsc{Hastie, T.} (2005). Regularization and
variable selection via the elastic net.
\textit{J. Roy. Statist. Soc. Ser. B}
\textbf{67} 301--320.
\MR{2137327}

\bibitem[\protect\citeauthoryear{}{2009}]{2009Zou}
\textsc{Zou, H.} and \textsc{Zhang, H. H.} (2009). On the adaptive
elastic net with a diverging number of parameters.
\textit{Ann. Statist.} \textbf{37} 1733--1751.
\MR{2533470}

\end{thebibliography}
\end{document}